\documentclass[aps,prl,floatfix,nofootinbib,superscriptaddress,twocolumn]{revtex4-1}
\usepackage{graphicx,subfigure}
\usepackage{amsmath,amssymb,amsfonts}
\usepackage{bm}
\usepackage{color}
\usepackage{comment}
\usepackage{slashed}
\def \be {\begin{equation} }
\def \ee {\end{equation}}

\def \bem {\begin{multline}}
\def \eem {\end{multline}}

\def \bes {\begin{subequations} }
\def \ees {\end{subequations}}

\newcommand{\Eq}[1]{Eq.(\ref{#1})}

\newcommand{\Fig}[1]{Fig.(\ref{#1})}

\begin{document}

\title{Multiparticle collectivity from initial state correlations\\ in high energy proton-nucleus collisions}
\author{Kevin Dusling}
\email{kdusling@mailaps.org}
\affiliation{American Physical Society, 1 Research Road, Ridge, New York 11961, USA}
\affiliation{Physics Department, Brookhaven National Laboratory, Bldg. 510A, Upton, New York  11973, USA}
\author{Mark Mace}
\email{mark.mace@stonybrook.edu}
\affiliation{Physics Department, Brookhaven National Laboratory, Building 510A, Upton, New York  11973, USA}
\affiliation{Physics and Astronomy Department, Stony Brook University, Stony Brook, New York 11974, USA}
\author{Raju Venugopalan}
\email{raju@bnl.gov}
\affiliation{Physics Department, Brookhaven National Laboratory, Bldg. 510A, Upton, New York 11973, USA}

\date{ \today}

\begin{abstract}
Qualitative features of multiparticle correlations in light-heavy ion ($p+A$) collisions at RHIC and LHC are reproduced in a simple initial state model of partons in the projectile coherently scattering off localized domains of color charge in the heavy nuclear target. These include i) the ordering of the magnitudes of the azimuthal angle $n$th Fourier harmonics of two-particle correlations $v_n\{2\}$, ii) the energy and transverse momentum dependence of the four-particle Fourier harmonic $v_2\{4\}$, and iii) the energy dependence of four-particle symmetric cumulants measuring correlations between different Fourier harmonics. Similar patterns are seen in an Abelian version of the model, where we observe $v_2\{2\} > v_2\{4\}\approx v_2\{6\}\approx v_2\{8\}$ of two, four, six, and eight particle correlations. While such patterns are often interpreted as signatures of collectivity arising from hydrodynamic flow, our results provide an alternative description of the multiparticle correlations seen in $p+A$ collisions.
\end{abstract}
\maketitle

A remarkable series of recent experiments at CERN's Large Hadron Collider (LHC) and at the Relativistic Heavy Ion Collider (RHIC) at BNL have demonstrated the existence of collective multiparticle dynamics in proton-proton ($p+p$) and light-heavy ion ($h+A$) collisions. Collectivity is represented by the behavior of $n$-th Fourier moments of the cumulants $c_n\left\{m\right\}$ of $m$-particle ($m\geq 4$) azimuthal angular anisotropy correlations; it is observed that corresponding real valued $m$-th roots of $c_n\left\{m\right\}$, the anisotropy coefficients $v_n\left\{m\right\}$, have nearly identical values for high multiplicity events. These results are similar to those obtained in peripherally overlapping collisions of heavy nuclei and even exhibit some of the systematics observed in fully overlapping central heavy-ion collisions. The collective dynamics of the latter is well described by sophisticated hydrodynamic models which presume the creation of a thermalized strongly interacting Quark-Gluon Plasma (QGP). 

Hydrodynamic models  have also been employed to describe the experimental results on multiparticle correlations in the smaller systems. Their agreement with data is however sensitive to the initial conditions for hydrodynamic flow~\cite{Schenke:2014zha}, specifically to the gluon ``shape" fluctuations within the proton~\cite{Mantysaari:2016ykx,Mantysaari:2017cni}. Since these shape fluctuations are themselves a consequence of strong initial state correlations, it is interesting to ask whether the collective properties of quark-gluon matter in the aforementioned small systems are those of nature's smallest fluids or whether there are alternative explanations for this collective behavior from initial state correlations alone. 

An initial state correlation scenario to explain the ``ridge-like" structure of two-particle correlations in small systems was advocated in \cite{Dumitru:2010iy} based on the  Color Glass Condensate (CGC) effective theory~\cite{Iancu:2003xm,Gelis:2010nm}. (Refs.~\cite{Kovner:2012jm,Dusling:2015gta,Schlichting:2016sqo} review these
and related frameworks.) For hadron transverse momenta $p_\perp,q_\perp\geq Q_{s,T}$, where $Q_{s,T}\sim 1-2$ GeV is the saturation scale of strongly correlated gluons in the nuclear target, a so-called Glasma graph scenario describes both ridge-like and jet-like correlations in the data on $p+p$ and $h+A$  collisions~\cite{Dumitru:2008wn,Dusling:2012iga,Dusling:2012cg,Dusling:2012wy,Dusling:2013qoz,Dusling:2015rja}. For lower momenta $p_\perp\leq Q_{s,T}$, corrections of order $Q_{s,T}/p_\perp$ are large. For two-particle correlations, the contribution of these corrections were quantified in \cite{Lappi:2015vta} and the qualitative behavior of the anisotropy coefficients $v_n\{2\}$ were reproduced\footnote{These computations are within a dilute-dense approximation $Q_{s,P} \ll Q_{s,T}$ in the CGC framework, where $Q_{s,P}$ is the projectile saturation scale. For high multiplicity events, corrections of order $Q_{s,P}/p_\perp$ are important and can be computed by numerically solving classical Yang-Mills equations~\cite{Schenke:2015aqa,Schenke:2016lrs} for two dense sources.}. 

Collectivity from four-particle initial state correlations has
remained elusive. Prior discussions were within a ``color domain"
model~\cite{Kovner:2010xk,Dumitru:2014dra,Dumitru:2014vka,Dumitru:2014yza,Skokov:2014tka} whose theoretical foundations are unclear~\cite{Lappi:2015vta}. In this Letter, we will
compute $v_2\left\{m\right\}$  for $m\geq 4$ systematically for the first
time in an initial state framework. An important ingredient is a first
computation of the average of the product of four light-like ``dipole" Wilson-line
correlators. We will explore the systematics of $v_2\left\{4\right\}$ 
as a function of $Q_{s,T}^2$, both
integrated and differential in $p_\perp$.  In addition, we will study so-called four-particle
symmetric cumulants which have recently been measured in
light-heavy ion collisions.  We observe strikingly that qualitative features of $v_n\{2\}$ and $v_2\{4\}$
measured in small systems are reproduced in this initial state framework. Since
the computational effort for $v_2\{m\}$ for $m > 4$ increases rapidly with $m$, the same computation can be carried out in an Abelian variant of the model.  Remarkably, the behavior of these higher cumulants are consistent with those observed in the LHC data. 

For our proof of principle computation\footnote{Further details of the computation are discussed in a longer paper~\cite{Dusling:2017aot}.}, we model the
proton-nucleus collision very simply as the eikonal scattering of nearly
collinear quarks in the projectile scattering off color domains of size $1/Q_{s,T}$ inside the
nuclear target~\cite{Dumitru:2002qt,Lappi:2015vha,Lappi:2015vta}. The $m$-particle correlation can be expressed as 
\begin{eqnarray}
\label{eqn:multiplicity}
&&\frac{d^m N}{d^2\mathbf{p_{i\perp}}\cdots d^2\mathbf{p_{m\perp}} }=
{\prod\limits_{i=1}^m} \int d^2\mathbf{b_i} \int
\frac{d^2\mathbf{k_i}}{(2\pi)^2} W_q(\mathbf{b_i},\mathbf{k_{i\perp}})
\nonumber\\
&&\cdot\int d^2\mathbf{r_i} e^{i(\mathbf{p_{i\perp}}-\mathbf{k_{i\perp}})\cdot \mathbf{r_i}} 
\left< \prod\limits_{j=1}^m D\left(\mathbf{b_j} +
\frac{\mathbf{r_j}}{2},\mathbf{b_j} - \frac{\mathbf{r_j}}{2}\right)\right>.
\end{eqnarray} 
Here we have made the simplifying assumption that the $m$-particle Wigner
function representing quark distributions in the incoming proton factorizes as
$W_{q}(\mathbf{b_1},\mathbf{k_1},...,\mathbf{b_m},\mathbf{k_m})=\prod\limits_{i=1}^m
W_{q}(\mathbf{b_i},\mathbf{k_i})$. Eikonal scattering is sensitive to the quark
dipole correlator $D(x,y)=\frac{1}{N_c}\text{Tr}\left[U(x)U^\dagger(y)\right]$, where 
$N_c$ is the number of colors and $U(x)$ ($U^\dagger(y)$) are light-like Wilson
lines appearing in the amplitude (complex conjugate amplitude) for
quarks multiple scattering off gluons in the target. In the
McLerran-Venugopalan (MV)
model~\cite{McLerran:1993ka,McLerran:1993ni,McLerran:1994vd}, these Wilson
lines are path ordered exponentials of color charges in the target, and the
average $\langle \cdots\rangle$ in \Eq{eqn:multiplicity} is performed over a
Gaussian distribution of color charges with a weight proportional to
$Q_{s,T}^2$~\cite{Iancu:2003xm,Lappi:2010ek}. We will assume further that the Wigner distributions of the nearly collinear quarks have
the Gaussian form
\begin{eqnarray}
W_q(\mathbf{b_i},\mathbf{k_{i\perp}})=\frac{1}{\pi^2}
e^{-|\mathbf{b_i}|^2/B_p}e^{-|\mathbf{k_i}|^2 B_p}\,,
\end{eqnarray}
where $B_p=4 ~\text{GeV}^{-2}$, a scale controlling the quark transverse momentum and spatial resolution, is fixed\footnote{In this model, $B_p$ also represents the transverse overlap area of the collision.} using dipole model fits to HERA data~\cite{Kowalski:2006hc,Rezaeian:2012ji}. We can perform the $\mathbf{k_i}$ integrals in \Eq{eqn:multiplicity} explicitly, which gives
\begin{eqnarray}
\label{eqn:multiplicity_kintegrated}
&&\frac{d^{m} N}{d^2\mathbf{p_{i\perp}}\cdots d^2\mathbf{p_{m\perp}} }=
\frac{1}{(4\pi^3 B_p)^m} {\prod\limits_{i=1}^m} \int d^2\mathbf{b_i}  \int
d^2\mathbf{r_i}~e^{-b_i^2/B_p}  \nonumber\\
&&~\cdot e^{-r_i^2/4B_p}  e^{i\mathbf{p_{i\perp}}\cdot \mathbf{r_i}}\left< \prod
\limits_{j=1}^m D\left(\mathbf{b_j} + \frac{\mathbf{r_j}}{2},\mathbf{b_j} -
\frac{\mathbf{r_j}}{2}\right)\right> .
\end{eqnarray} 

Before we proceed to the computation of multiparticle cumulants, we will address some of the features and limitations of this simple model. First, we note that even though rapidity is not explicit in this model, particle correlations are ridge-like and long range in rapidity. As we demonstrate explicitly in~\cite{Dusling:2017aot}, these correlations can be obtained in our model by convoluting the longitudinal momentum distributions of the quarks with their parton distributions in the incoming proton. Our model shares these features with the hybrid framework of multiparticle correlations discussed in \cite{Dumitru:2002qt,Kovner:2012jm,Kovchegov:2012nd}. This hybrid scenario will receive significant modifications when high parton density effects in the projectile become important. These effects quantitatively go as $Q_{s,P}^2/p_\perp^2$; saturation models fit to HERA data conservatively suggest that these effects become non-negligible around $x=0.01$~\cite{Rezaeian:2012ji}. However, depending on the transverse momentum range studied, the qualitative features we observe could persist to smaller values of $x$. Parametrically the rapidity range where corrections to the hybrid model occur is $\Delta y \gtrsim 1/\alpha_S$.

%These effects quantitatively go as $Q_{s,p}^2/p_T^2$. Saturation models fit to HERA data indicate that $Q_{s,p} << p_T$, where $p_T \geq 1$ GeV, even at $x_q=0.001$. Qualitative estimates suggest that {\it inclusive} correlations between valence quarks or valence quarks and sea quarks ($x_1$ and $x_2$ for our purposes) may receive significant corrections when gluon emission is copious enough to generate large parton densities; this would occur parametrically for $\alpha_S\ln(x_1/x_2)\geq1$, or $\Delta y \geq 1/\alpha_S$. 

Second, an obvious limitation of our model is that it only includes quarks. This is clearly not sufficient at the highest RHIC energies and at the LHC, though it may suffice to explain the ridge like correlations now seen at fairly low energies in deuteron-gold collisions at RHIC. Our model can be extended to include gluon degrees of freedom from the projectile; the only modification is that they will be color rotated by adjoint Wilson lines from the target and one has to compute color traces of products of these Wilson lines instead. However such computations alone are insufficient because they do not generate odd moments of the azimuthal distributions, a consequence of the generators of the adjoint representation being real~\cite{Kovchegov:2012nd,Dusling:2013qoz,Kovchegov:2013ewa}. The odd moments can, however, be recovered by going beyond the strict dilute-dense limit and including gluon exchanges between spectator partons and the scattered gluons in the projectile~\cite{Schenke:2015aqa,McLerran:2016snu,Kovner:2016jfp}. These considerations are at present beyond the scope of this Letter. 

The two- and four-particle cumulants are defined as~\cite{Borghini:2001vi} 
\begin{eqnarray}
c_n\left\{2\right\}&=&\left< e^{in(\phi_1-\phi_2)} \right> \equiv \frac{\kappa_{n}\left\{2\right\}}{\kappa_{0}\left\{2\right\}}\,,
\label{eqn:cumulant_cn2}
\end{eqnarray}
and
\begin{eqnarray}
c_n\left\{4\right\}&=&\left< e^{in(\phi_1+\phi_2-\phi_3-\phi_4)} \right> -2
    \left< e^{in(\phi_1-\phi_2)} \right>^2 \nonumber \\
&\equiv&  \frac{\kappa_{n}\left\{4\right\}}{\kappa_{0}\left\{4\right\}}-2\left(\frac{\kappa_{n}\left\{2\right\}}{\kappa_{0}\left\{2\right\}}\right)^2\,,
\label{eqn:cumulant_cn4}
\end{eqnarray}
with
\begin{eqnarray}
\label{eqn:cumulant_multiplicity}
\kappa_{n}\{2\} &=& \int \frac{d^{2} N}{
d^2\mathbf{p}_{1\perp}d^2\mathbf{p}_{2\perp}}  e^{ in(\phi^{p}_1-\phi^{p}_2)} \\
\kappa_{n}\{4\} &=&
\int
\frac{d^{4}
N}{d^2\mathbf{p}_{1\perp}d^2\mathbf{p}_{2\perp}d^2\mathbf{p}_{3\perp}d^2\mathbf{p}_{4\perp}}
e^{ in(\phi^{p}_1+\phi^{p}_2-\phi^{p}_3-\phi^{p}_4)}\nonumber\,,
\end{eqnarray}
where the integration over the two and four-particle phase space is implicit.

The computation of two-particle cumulants is straightforward. The corresponding anisotropy coefficients are defined to be~\cite{Borghini:2001vi}
\begin{eqnarray}
v_n\{2\}=(c_n\{2\})^{1/2} \,,
\label{eqn:vn2}
\end{eqnarray}
and were computed previously for the MV model in \cite{Lappi:2015vta}. In computing these, we first fix the value of $Q_{s,T}$ and integrate over the momenta of all $m$ particles in the range $p_\perp\in [0,p_\perp^{\rm max}]$. We then study the variation of $v_n\{2\}$ with increasing $Q_{s,T}^2$; in our simple model, this corresponds to increasing the center-of-mass energy or the centrality of the collision. By construction, our results are independent of the number of charged particles, $N_{\rm ch}$, produced in the collision.
\begin{figure}
\includegraphics[width=.5\textwidth]{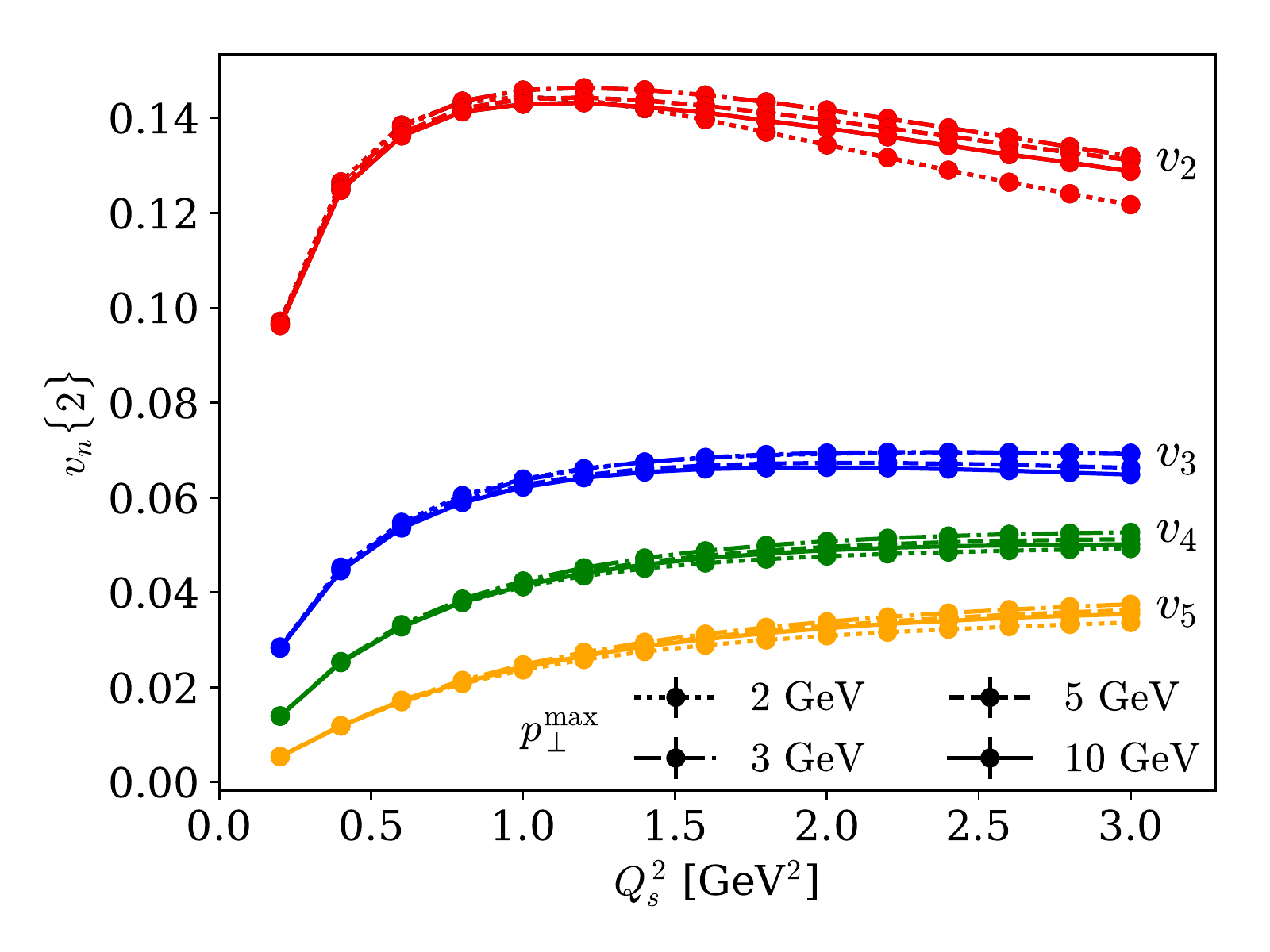}
\caption{Two-particle Fourier harmonics defined in \Eq{eqn:vn2}, as a function of harmonic $n$ and $Q_{s,T}^2$, for four different values of the maximum integrated transverse momentum, $p_\perp^{\rm max}$.}\label{fig:vn2}%
\end{figure}
In \Fig{fig:vn2}, we plot the two-particle Fourier harmonics for $n=2,..,5$ as a function of $Q_{s,T}^2$.
The upper limit of the transverse momentum integration is taken for multiple values to ensure convergence; by $p_\perp^{\rm max}=3~\text{GeV}$ the results are no longer dependent on $p_\perp^{\rm max}$. We observe a clear ordering of the $n$ harmonics. These observations are in qualitative agreement with experiment. We should caution, however, that there is not a simple one-to-one map between $Q_{s,T}$ and the energy or centrality. Further, as noted in \cite{Lappi:2015vta}, the QCD evolution of the MV model with energy will lead to lower values of $v_n\{2\}$. Fragmentation of gluons into hadrons will further soften the signal~\cite{Schenke:2016lrs}. Our results therefore represent maximal values  for azimuthal correlations in this initial state framework. 

We will now go beyond the study in \cite{Lappi:2015vta} and discuss the behavior of the four-particle flow coefficient defined as~\cite{Borghini:2001vi}
\begin{eqnarray}
v_n\{4\}=(-c_n\{4\})^{1/4}\,.
\label{eqn:vn4}
\end{eqnarray}
The computation of four-particle cumulants is significantly more complex for two reasons. First, as is clear from Eqs.(\ref{eqn:multiplicity_kintegrated})-(\ref{eqn:cumulant_multiplicity}), one has 24 integrals to perform relative to 12 previously for two-particle cumulants.
However, more importantly, computing the expectation value of the product of four dipole correlators is non-trivial. While an analytical expression exists for the expectation value of the product of two dipoles~\cite{Lappi:2015vta,Blaizot:2004wu,Dominguez:2008aa}, such an expression is not known for the product of four dipoles. For work in this direction, see~\cite{Fukushima:2008ya}. 

Our strategy for computing $n$-dipole correlators follows the framework
introduced in \cite{Blaizot:2004wu}. For four dipoles, the building blocks are
four light-like Wilson lines, localized at distinct transverse positions, in the scattering amplitude for four
quarks along the light cone from $x^+=-\infty$ to $x^+=+\infty$ and their counterparts in the complex conjugate amplitude. The terms in the expansion of the Wilson lines correspond to multiple gluon exchanges, ordered in the $x^+$ direction, between the dipoles. These exchanges generate quadrupole, sextupole, and octupole configurations that are, respectively, the traces over the product of four, six, and eight light-like Wilson lines\footnote{For a number of dilute-dense multiparticle processes only dipoles and quadrupoles contribute in the high energy limit~\cite{Dominguez:2012ad}. However because the leading contributions to the correlation observables here are themselves $N_c$ suppressed, sextupoles and octupoles are of equal importance.}. Permutations of the coordinates for each of these topologies results in 24 distinct basis elements. Subsequent gluon exchanges generate transitions between elements of this basis,  corresponding to a $24\times 24$ matrix that can be exponentiated numerically. 

All the basis elements are known in the MV model and the result of the computation can be expressed in terms of the saturation scale $Q_{s,T}$ and a cutoff $\Lambda$ regulating the infrared behavior of the two-dimensional gluon exchange propagator. We choose $\Lambda= 0.241$ GeV~\cite{Lappi:2013zma}; our results are insensitive to variations in this scale\footnote{For values $Q_{s,T}^2 \approx (p_\perp^{\rm max})^2  \gg\Lambda^2$, the results are insensitive to $Q_{s,T}^2 B_p$, the number of color domains in the target~\cite{Dusling:2017aot}.}. The procedure can be extended to the average of $m>4$ dipole correlators and is discussed at length in \cite{Dusling:2017aot}.

\begin{figure}
\includegraphics[width=.5\textwidth]{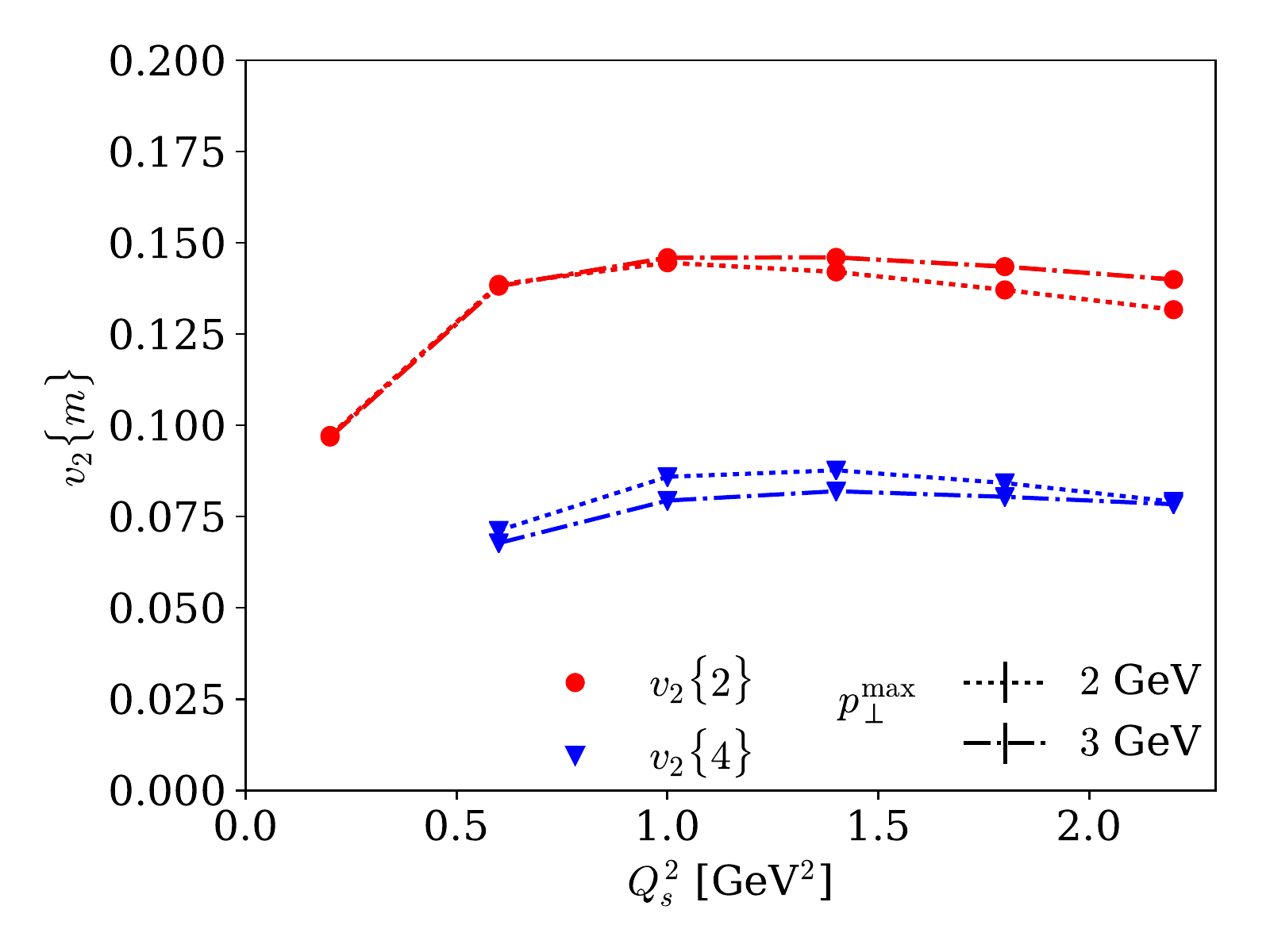}
\caption{Two and four-particle Fourier harmonics defined in \Eq{eqn:vn2} and \Eq{eqn:vn4} respectively, as a function of $Q_{s,T}^2$, for two different values of the maximum integrated transverse momentum, $p_\perp^{\rm max}$.}
\label{fig:v2int}%
\end{figure}

Computing \Eq{eqn:multiplicity} as outlined, we can evaluate $v_n\{4\}$ using Eqs.~(\ref{eqn:cumulant_cn4}), (\ref{eqn:cumulant_multiplicity}), and (\ref{eqn:vn4}). The results are shown in \Fig{fig:v2int}. For reference, we also plot the $v_2\{2\}$ values shown in \Fig{fig:vn2} for $p_\perp^{\rm max}=2, 3$ GeV. 
First, we see striking evidence of collectivity as defined: the value of $c_2\{4\}$ is negative allowing us to extract a real valued $v_2\{4\}$. It is smaller than the value of $v_2\{2\}$ and both of these are relatively flat as a function of $Q_{s,T}^2$. Since the increase in $Q_{s,T}$ corresponds to an increase in the center-of-mass energy, the two and four-particle elliptic anisotropy coefficients are independent of energy in our initial state model. Experimental results from light-heavy ion collisions at RHIC and LHC similarly show a weak variation of these quantities across a very wide window of center-of-mass energies~\cite{Aad:2013fja,Chatrchyan:2013nka,Abelev:2014mda,Belmont:2017lse,Aaboud:2017acw}. As noted previously, our results are independent of $N_{\rm ch}$, as is also approximately the case in experiment. 

To compute the two and four-particle Fourier harmonics for a fixed transverse momentum, we define 
\begin{eqnarray}
d_n\{2\}(p_\perp)  \equiv \frac{\tilde{\kappa}_{n}\left\{2\right\}(p_\perp)}{\tilde{\kappa}_{0}\left\{2\right\}(p_\perp)}\,,
\label{eqn:cumulant_dn2}
\end{eqnarray}
and
\begin{eqnarray}
d_n\{4\}(p_\perp) \equiv \frac{\tilde{\kappa}_{n}\{4\}(p_\perp)}{\tilde{\kappa}_{0}\{4\}(p_\perp)}-2\frac{\tilde{\kappa}_{n}\{2\}(p_\perp) ~\kappa_{n}\{2\} }{\tilde{\kappa}_{0}\{2\}(p_\perp) ~\kappa_{0}\{2\}} \, .
\label{eqn:cumulant_dn4}
\end{eqnarray}
The ${\tilde \kappa}_n$ are the differential form of $\kappa$ where one of the momenta are not integrated over. The $v_n$ anisotropies as a function of $p_\perp$ are given by~\cite{Aad:2013fja}
\begin{eqnarray}
v_n\{2\}(p_\perp)=\frac{d_n\{2\}(p_\perp)}{(c_n\{2\})^{1/2}} \,\,;\,\, v_n\{4\}(p_\perp)=\frac{-d_n\{4\}(p_\perp)}{(-c_n\{4\})^{3/4}}\, . \nonumber \\
\label{eqn:vnpt}
\end{eqnarray}
\begin{figure}
\includegraphics[width=.5\textwidth]{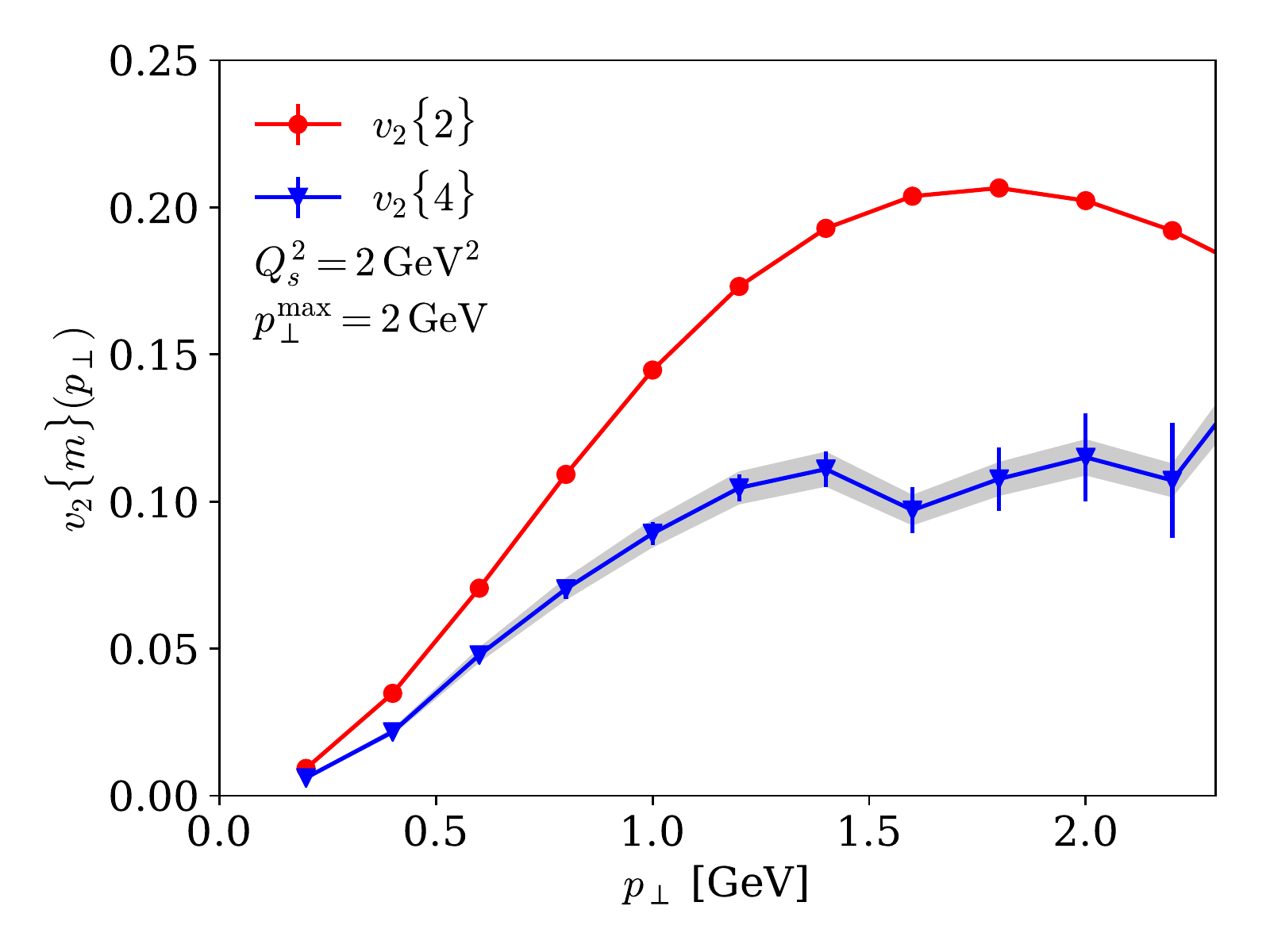}
\caption{The $v_n\{m\}$ Fourier harmonics defined in \Eq{eqn:vnpt} as a function of $p_\perp$ for fixed $Q_{s,T}^2$. The $m-1$ momenta are integrated from $p^{int}_\perp=0$ to $2~\text{GeV}$.}
\label{fig:vnpt}%
\end{figure}
We plot these for $n=2$ and $m=2, 4$ in \Fig{fig:vnpt} for a fixed $Q_{s,T}^2= 2$ GeV$^2$. The error bands represent the systematic uncertainty in the integrated four-particle cumulant. We again see that $v_2\{m\}(p_\perp)$ has the same qualitative behavior as data in light-heavy ion experiments at RHIC and LHC.  Energy evolution of parton distributions and parton to hadron fragmentation will decrease the values shown. Increasing $Q_{s,T}^2$ has the effect of flattening out $v_n\{m\}(p_\perp)$ at higher $p_\perp$~\cite{Dusling:2017aot}.

Symmetric cumulants defined as 
\begin{eqnarray}
{\rm SC}(n,n')=\frac{{\bar \kappa}_{n,n'}\{4\}}{\kappa_{0}\{4\}}-\frac{\kappa_{n}\{2\}\kappa_{n'}\{2\}}{\kappa_{0}\{2\}^2} 
\label{eqn:sym_cumulant}
\end{eqnarray}
have recently been measured at the LHC. The ${\bar \kappa}_{n,n'}$ are analogous to $\kappa_n$ in \Eq{eqn:cumulant_multiplicity}, except that the odd numbered azimuthal angles have harmonic $n$, whereas the even angles have the harmonic $n'$. The $\text{SC}(n,n')$ cumulants directly measure correlations between the different flow harmonics~\cite{Bilandzic:2013kga}. In heavy-ion collisions, the data show that $\text{SC}(2,3)$ are increasingly anti-correlated with increasing centrality percentile, while the $\text{SC}(2,4)$ cumulant are increasingly correlated; these systematics are also seen in hydrodynamic models~\cite{Zhu:2016puf}.

\begin{figure}
\includegraphics[width=.5\textwidth]{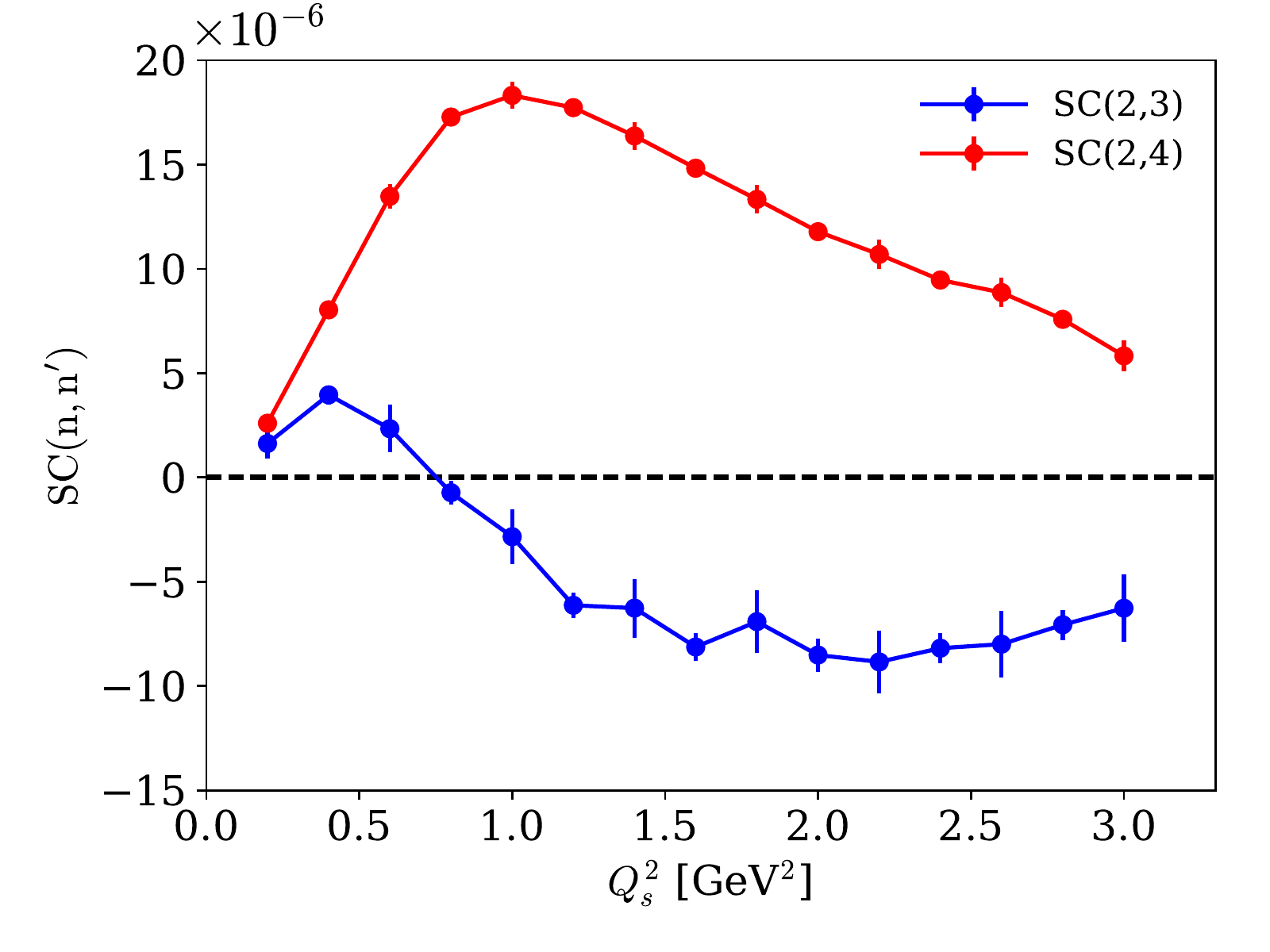}
\caption{Four-particle symmetric cumulants defined in \Eq{eqn:sym_cumulant}, as a function of $Q_{s,T}^2$.}
\label{fig:SC2434}
\end{figure}

The $\text{SC}(2,3)$ and $\text{SC}(2,4)$ cumulants have also been measured in light-heavy ion collisions and show the same pattern of correlations as for heavier systems~\cite{CMS:2017saf}. This is perhaps not too surprising because the correlations/anti-correlations are most significant in peripheral heavy-ion collisions. Our results for $\text{SC}(2,3)$ and $\text{SC}(2,4)$ are shown in \Fig{fig:SC2434} as a function of $Q_{s,T}^2$. We see that $\text{SC}(2,3)$ is negative by $Q_{s,T}=1$ GeV while $\text{SC}(2,4)$ is positive for all $Q_{s,T}$. Our results demonstrate clearly that such patterns are not unique to an interpretation requiring hydrodynamic flow. Results from hydrodynamic computations for these cumulants in light-heavy ion collisions are not yet available. 

To gain further insight into our results, it is useful to ask whether coherent multiple scattering off the target is crucial. One way to test this within our framework is to employ the Glasma graph approximation~\cite{Dumitru:2008wn,Dumitru:2010iy,Dusling:2012iga,Dusling:2012cg,Dusling:2012wy,Dusling:2013qoz,Dusling:2015rja},  valid for $p_\perp > Q_{s,T}$. For two partons scattering off the target color fields, the Glasma graph approximation corresponds to two gluon exchange in the scattering amplitude~\cite{Lappi:2015vta}. Non-linearities, that are large for $p_\perp \leq Q_{s,T}$ in the MV model, arise from multiple gluon exchanges between the projectile and the target. 
One can similarly implement the ``linear" Glasma graph approximation for four partons scattering off the target; we find $c_2\{4\}$ for the Glasma graphs is positive~\cite{Dusling:2017aot}--this confirms the importance of coherent multiple scattering. 

\begin{figure}
 \includegraphics[width=.5\textwidth]{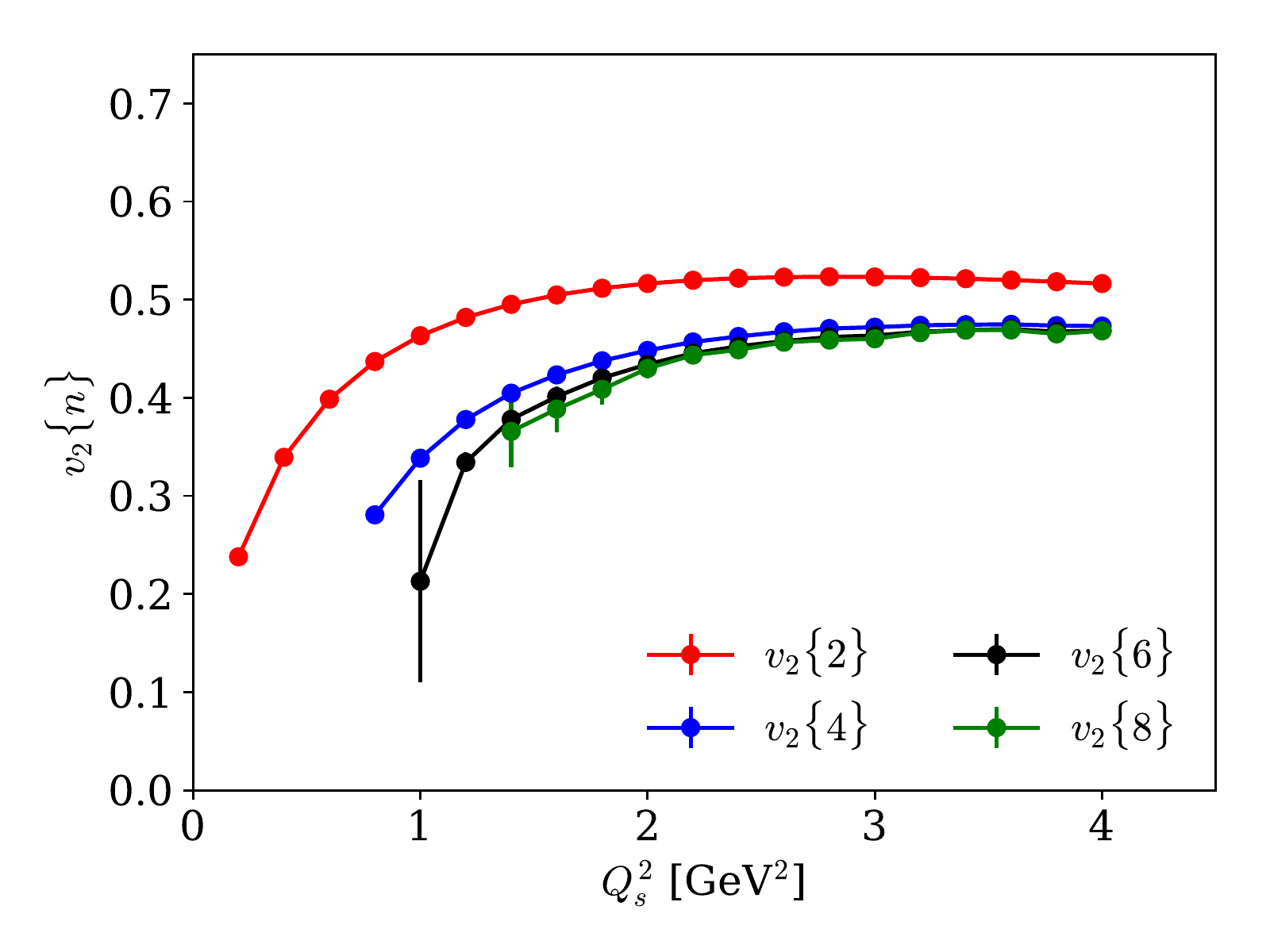}
\caption{Two-, four-, six- and eight-particle Fourier harmonics for coherent multiple scattering off Abelian fields plotted as a function of $Q_{s,T}^2$.}
\label{fig:Abelian}
\end{figure}

It is also interesting to consider coherent multiple scattering in the Abelian limit of this model. In this case, the Wilson lines are not matrices
in color space, but simply path ordered exponentials~\cite{Bjorken:1970ah}. The
product of dipoles in \Eq{eqn:multiplicity} is significantly simpler to
compute~\cite{Dusling:2017aot}, enabling one to extract $v_2\{6\}$ and
$v_2\{8\}$ from the corresponding cumulants~\cite{Borghini:2001vi,Khachatryan:2015waa}. Our
results, shown in \Fig{fig:Abelian}, demonstrate that $v_2\{2\} > v_2\{4\} \approx
v_2\{6\}\approx v_2\{8\}$, as also seen in the LHC data on multiparticle harmonics~\cite{Khachatryan:2015waa,Aaboud:2017acw}.

The fact that this behavior is reproduced in a simple initial state model is a proof of principle that it is not unique to interpretations of collectivity arising from the hydrodynamic response of the system to the $n$-th moments of $m$ particle spatial eccentricities~\cite{Bozek:2012gr,Kozlov:2014fqa,Bzdak:2013rya,Giacalone:2017uqx}. For a recent review on hydrodynamic collectivity and relevant references, see~\cite{Song:2017wtw}. Our results do not necessarily mean that an initial state interpretation of the data is favored. We instead conclude that the $v_n\{m\}$ measurements alone are insufficient to unambiguously distinguish between initial and final state approaches. 

While it is remarkable that our results qualitatively explain observed multiparticle correlations, it is also clear that the model is missing key features of QCD dynamics that should be important at high energies. In this regard, the initial state framework in~\cite{Schenke:2015aqa,Schenke:2016lrs} includes a more systematic treatment, albeit at an enormously greater computational effort. Nevertheless, since multiparticle correlations display similar features in light-heavy ion collisions spanning two orders of magnitude in center-of-mass energies, where QCD degrees of freedom evolve significantly, it is worth thinking further why this simple model appears to capture the underlying dynamics.

\begin{acknowledgments}
We would like to thank Jiangyong Jia, Tuomas Lappi, Jean-Fran\c{c}ois Paquet, Bj\"{o}rn Schenke, S\"{o}ren Schlichting, Chun Shen, Vladimir Skokov, and Prithwish Tribedy for useful discussions. R.V. would like to thank J\"{u}rgen Schukraft for motivational remarks. This material is based on work supported by the U.S. Department of Energy, Office of Science, Office of Nuclear Physics, under Contracts No. DE-SC0012704 (M.M.,R.V.) and DE-FG02-88ER40388 (M.M.). M.M. would also like to thank the BEST Collaboration for support. This research used resources of the National Energy Research Scientific Computing Center, a DOE Office of Science User Facility supported by the Office of Science of the U.S. Department of Energy under Contract No. DE-AC02-05CH11231 and the LIRED computing system at the Institute for Advanced Computational Science at Stony Brook University. 
\end{acknowledgments}

\end{document}